**Broadband Dual-Comb Spectroscopy in the Spectral Fingerprint Region**

O. Kara,[1] L. Maidment,[1] T. Gardiner,[2] P. G. Schunemann,[3] and D. T. Reid[1]

1. Scottish Universities Physics Alliance (SUPA), Institute of Photonics and Quantum Sciences, School of Engineering and Physical Sciences, Heriot-Watt University, Edinburgh EH14 4AS, UK.

2. National Physical Laboratory, Hampton Road, Teddington, London TW11 0LW, UK.

3. BAE Systems, Inc., MER15-1813, P.O. Box 868, Nashua, NH, USA 03061-0868

Author e-mail address: ok45@hw.ac.uk

Infrared spectroscopy in the spectral fingerprint region from 6–12 µm accesses the largest molecular absorption cross-sections, permitting sensitive, quantitative and species-specific measurements. Here, we show how dual-comb spectroscopy—a form of high-speed Fourier-transform spectroscopy involving no moving parts and capable of very high resolutions—can be extended to the 6–8-µm wavelength band using femtosecond optical parametric oscillators (OPOs). By acquiring dual-comb interferograms faster than the mutual decoherence time of the OPO combs we implement line-shape-preserving averaging to obtain low-noise, high-fidelity spectra of $H_2O$ and $CH_4$ at approximately 0.3-$cm^{-1}$ resolutions from 1285–1585 $cm^{-1}$.

_______________________________________

For several decades, Fourier-transform spectroscopy (FTS) has been regarded as a gold-standard analytic technique, with laboratory spectrometers incorporating broadband thermal sources covering the infrared region, in which different molecules and even molecular isomers can be distinguished by their unique spectral absorption features. Dual-comb spectroscopy (DCS) [1, 2] is a laser-based analog to FTS, permitting rapid, high-resolution measurements in a system containing no moving parts. Implementations of DCS in the near-infrared have exploited the excellent spatial coherence of laser illumination to perform spectroscopy over kilometer paths, enabling the sensitive detection of trace gases and pollutants [3]. To date, most DCS systems have been reported in the near-infrared (near-IR) [4-6] because of the maturity of frequency-comb sources in this region [7-9]. The extension of DCS to the mid-IR has been achieved using difference-frequency generation (DFG) sources [10-13], quantum cascade lasers [14-16], micro-resonators [17], singly-resonant optical parametric oscillators (OPOs) [18-21] and doubly-resonant OPOs [22]. However, DCS in the mid-IR fingerprint region has been limited to an early demonstration using a µW-level DFG source, which achieved 2-

cm$^{-1}$ resolutions from 9–12 µm [10], and with a quantum-cascade laser comb, whose wide mode spacing required a multi-hour sweep of the comb-line positions in order to resolve molecular line-shapes over a 15-cm$^{-1}$ range near 1420 cm$^{-1}$ [16]. Indeed, quantum-cascade lasers are currently the only means of achieving moderately broadband high-resolution spectroscopy above 6 µm, with the technique of multi-heterodyne spectroscopy showing potential for coverage of tens of cm$^{-1}$ [23] but with actual reported absorption spectroscopy only spanning < 1 cm$^{-1}$ near 1190 cm$^{-1}$ [23, 24]. Broadband multi-heterodyne spectroscopy also suffers from discontinuities in its spectral coverage, which arise from regions of the spectrum where the multi-heterodyne signal is weak, because FM modulation redistributes the power unevenly among the laser modes. Furthermore, properly resolving narrow line shapes requires scanning of the comb modes, which is time consuming and sacrifices the temporal resolution of the technique. These limitations can be circumvented by using new semiconductor nonlinear crystals to extend the wavelength coverage of near-IR femtosecond lasers to a region spanning from 6–12 µm [25]. This wavelength range is commonly known as the molecular 'fingerprint' region, where a wide range of organic compounds can be discriminated through the complex set of absorption features they exhibit due to molecular bending transitions. Here we introduce a dual-comb source for the spectral fingerprint region based on a pair of entirely free-running OPOs, each pumped by a 1-µm femtosecond laser and utilizing the new gain medium orientation-patterned gallium phosphide (OPGaP) to produce broadband idler pulses tunable from 6–8 µm.

**Results**

Dual-comb spectroscopy using OPOs has been demonstrated in the 3- to 4-µm region using 1-µm-pumped MgO:PPLN singly-resonant OPOs [19–21], and up to 5.3 µm using 1.93-µm-pumped OPGaAs degenerate doubly-resonant OPOs [22]. Obtaining high quality spectroscopy has typically required fully-locked combs [11], which provide superior performance to free-running systems, in which the carrier frequency of the DCS interferogram shifts as the comb offsets fluctuate (for example, [10]). The results we present here demonstrate that such fluctuations need not present a barrier to high-fidelity dual-comb spectroscopy with OPOs, so long as each interferogram is recorded on a timescale faster than the characteristic decoherence time between the OPOs, i.e. the time in which their comb-mode frequencies change significantly. Our DCS system operates with free-running 102-MHz modelocked OPOs, whose difference in pulse repetition frequencies ($\Delta f_{REP}$) exceeds 1 kHz, implying that high resolution spectroscopy can be obtained from interferograms acquired in a few tens of microseconds.

The DCS system (see Methods) is presented in Fig. 1, and comprises two OPGaP OPOs resonant near 1.2 µm and generating idler pulses tunable from 6–8 µm, depending on the grating period used. Spectroscopy is implemented in a symmetric (homodyne) configuration, in which pulses from both combs interact with the spectroscopic sample. Each interferogram is recorded, Fourier transformed and provided with a relative optical frequency calibration obtained from concurrent measurements of $\Delta f_{REP}$ and $f_{REP}$. A correlation-based co-alignment algorithm (see Fig. 2 and Methods) is used to correct for the systematic frequency shifts between all the members of a dataset containing more than one hundred consecutively acquired spectra, following which simple averaging provides a single high-quality spectrum.

**Spectroscopic measurements of ambient water vapor** Operating on a grating period of 27 µm, the OPOs produce pulses with spectra spanning 1280–1370 cm$^{-1}$ and which travel 2.5 m in air from the OPGaP crystals to the detector. Water vapor absorption is relatively weak in this region but can still be clearly seen in the spectrum shown in Fig. 3a. Envelope removal and fitting to the HITRAN 2012 database yields the transmission spectrum in Fig. 3d. The regular modulations evident in the residuals of the trace (Fig. 3g, blue) are étalon effects which can be directly traced to parasitic reflections in the OPGaP crystals, a Ge filter and the ZnSe idler output couplers in the OPO cavities. The transmission spectrum can be corrected by removing these étalon effects, and doing this results in excellent agreement with HITRAN at 0.256-cm$^{-1}$ resolution, with features and line-shapes becoming clearly resolved (Fig. 3j). The remaining residual (Fig. 3g, green) shows a standard deviation of 1.3%, implying a noise-equivalent absorption of 0.0041 Hz$^{-1/2}$. The fit to HITRAN allows quantitative determination of the water vapor concentration, yielding a mole-fraction concentration of 0.54%, consistent with our lab conditions.

With a grating period of 24.5 µm the OPO produces shorter idler wavelengths which span the 1490–1590-cm$^{-1}$ spectral region. Performing an identical analysis leads to the results in Fig. 3b which again show excellent agreement with the HITRAN simulation in this region (Fig. 3e,k). The impact of étalons is much less severe in this band because of the better performance of the long-wave IR anti-reflection coatings on our optics at these wavelengths. The fitting resolution in this case is 0.298-cm$^{-1}$ and the mole-fraction water-vapor concentration was determined as 0.77%. The difference in ambient water-vapor concentration compared to the earlier result is not unexpected, since the lab humidity varies with time and with local atmospheric conditions. After étalon-fringe removal the residual (Fig. 3h, green) shows a standard deviation of 2.4%, giving a noise-equivalent absorption of 0.0076 Hz$^{-1/2}$.

**Combined spectroscopic measurements of methane and ambient water vapor** With the OPO operating in the 1280–1370-cm$^{-1}$ band, dual-comb interferograms are acquired after introducing into both OPO beams a 20-cm gas cell containing a synthetic air mixture with nominally 2% methane. Processing the interferograms as previously described yields the results in Fig. 3c,f. As the distance from the OPGaP crystals to the detector remains the same as before, a similar water-vapor contribution to the spectrum is both expected and observed. After étalon removal (Fig. 3i, green) the standard deviation of the residuals of 3.5% implies a noise-equivalent absorption for methane at this concentration of 0.011 Hz$^{-1/2}$. Our fitting procedure simultaneously and independently fits for methane and water-vapor concentrations and yields parameters of 0.68% water vapor and 2.17% methane, consistent with previous ambient air measurements and with the original filling specification of the methane gas cell. Excellent agreement is obtained (Fig. 3l) with the HITRAN data with a fitting resolution of 0.314 cm$^{-1}$.

**Discussion**

The results reported here are the first examples of combined broadband and high-resolution spectroscopy in the spectral fingerprint region from 6–8 µm. They also represent among the highest quality dual-comb data obtained in the mid-IR using free-running femtosecond lasers, demonstrated by the excellent agreement between the HITRAN 2012 database and the positions, relative magnitudes and shapes of the experimentally measured absorption lines. This performance has been achieved by employing a novel line-shape-preserving averaging protocol, which we demonstrated for approximately 115 spectra but that in principle could be applied to a much larger dataset without loss of fidelity. Unlike other line-by-line fitting approaches, the availability of a single broadband spectrum allows us to conduct a full-spectrum multi-parameter fit to the HITRAN database, which our results show is robust in the presence of noise and of étalon effects, allowing these to be eliminated by post-processing. The robustness of the fit in the presence of étalon effects was confirmed by re-running the fit after removing the étalon fringes, which led to no significant change in the best-fit parameters.

The interferogram time window used for data processing is 40 µs, implying a limiting resolution of 0.076 cm$^{-1}$, a factor of 3–4 lower than the experimentally observed resolutions of 0.256–0.314 cm$^{-1}$. We believe this difference is caused by decoherence between the combs that occurs on the timescale of the interferogram window. Acquiring interferograms in a shorter time by operating the system at a higher value of $\Delta f_{REP}$ should therefore mitigate this problem, although steps would need to be taken to limit the optical bandwidth in order to avoid aliasing.

The implications of extending DCS into the spectral fingerprint band and in a manner not requiring stabilization of either $f_{REP}$ or $f_{CEO}$ are far reaching. In addition to the organic fingerprints, many pollutant gases also exhibit high absorption cross sections in this region (e.g. $NH_3$, $CH_4$, $N_2O$), meaning that sensitive and quantitative free-space detection for the purposes of environmental monitoring is possible, without the need for complex phase-locked systems. The potential of DCS for multi-kHz data rates means that chemical reactions could be monitored directly in real time, and the specificity of the fingerprint region to molecular isomers promises to provide new insights into the dynamics of chemical bonding [26]. Similar advantages apply to nano-Fourier transform infrared spectroscopy (FTIR) [27], in which scanning near-field optical microscopy is implemented with mid-IR illumination of a metallic tip, enabling nanoscale chemical mapping of polymer blends, organic fibers, and biomedical tissue. Dual-comb illumination could dramatically increase the acquisition rate of nano-FTIR spectra, potentially even enabling a form of nanoscale hyperspectral imaging.

**Acknowledgements**

This work was financially supported by the UK Engineering and Physical Sciences Research Council and the UK National Physical Laboratory under an Industrial CASE award. (Ref. 13220137).

**Author contributions**

LM and OK designed and constructed the OPGaP OPOs. OK developed the DCS set-up and carried out the measurements. DTR conceived the experiments and led the data analysis, with contributions from TG. All authors contributed to the content of the paper.

**Methods**

**Configuration of the OPGaP OPOs and DCS system**

The DCS system (Fig. 1a) comprises two OPGaP OPOs which follow a design similar to that reported in [25]. Each OPO is pumped by a femtosecond Yb:fiber laser whose pulse repetition frequency (nominally 102.0 MHz) can be adjusted using an intracavity piezoelectric actuator. Before entering the OPO, the pulses from each pump laser are amplified and then compressed to durations close to their bandwidth-limited values of 150 fs. The OPGaP OPOs are operated on grating periods of 24.5 µm to obtain signal (idler) wavelengths centered around 1.25 µm (6.5 µm) and 27 µm for signal (idler) wavelengths centered near 1.22 µm (7.5 µm). The instantaneous bandwidth of the idler pulses in each case is approximately 500 nm. The idler pulses are extracted from each OPO using a silver intracavity curved mirror situated immediately after each OPGaP crystal (Fig. 1b) and an idler-transmitting / signal-reflecting coating on a plane

ZnSe intracavity mirror. This arrangement yields well collimated idler beams which are combined at a $CaF_2$ beamsplitter whose coating is approximately 50% reflecting from 2–8 μm. Dual-comb interferograms (Fig. 1c) are measured from both channels of the beamsplitter using $LN_2$-cooled HgCdTe detectors with a bandwidth of 50 MHz (Kolmar KMPV11-0.25-J1) . Up to 120 interferograms (with this number limited by the buffer size of the acquisition unit) are recorded in a single acquisition (Fig. 1c), following which each interferogram is automatically centered (Fig. 1d) and Fourier-transformed to yield its radio-frequency spectrum (Fig. 1e). Sample-gas spectroscopy, in this case methane, is conducted by inserting a 20-cm gas cell before one of the detectors, while background spectroscopy, in this case water vapor, is carried out with the other detector. Both interferograms are digitized at 200 MHz and acquired on two channels of a 14-bit-resolution USB oscilloscope. A third channel records a signal at $10 \Delta f_{REP}$, which is obtained by isolating the tenth harmonic of each pump laser's repetition frequency (RF) using a narrowband 1-GHz bandpass filter, then differencing the two repetition-frequency harmonics using a RF mixer followed by a low-pass filter. A zero-crossings algorithm accurately extracts $\Delta f_{REP}$ for each interferogram, following which the corresponding RF spectrum is then scaled to the optical domain by multiplying its frequency sampling interval by a factor of $f_{REP} / \Delta f_{REP}$ . Finally, each spectrum obtained in this way is recorded as one row in a matrix, in preparation for subsequent spectral co-alignment and averaging with other spectra.

**Spectral co-alignment and averaging**

Fluctuations in $f_{CEO}$ for each of the two combs occur on millisecond timescales (the mutual decoherence times of the combs), resulting in small variations in the RF carrier frequency of each dual-comb interferogram. So long as each interferogram is recorded in a time much faster than this decoherence time then the spectral information remains intact but is simply shifted to an alternative center frequency. Spectral averaging can therefore be performed to improve the signal:noise ratio of the data, but before this can be done, each spectrum must be accurately co-aligned with every other spectrum. We implement this using a full-spectrum cross-correlation technique as follows.

Consider $N$ similar spectra, each of which is centered at a different position in the sampling grid. We calculate the relative shift, $\Delta \nu$, between any pair of spectra, $I_i(\nu)$ and $I_j(\nu)$ ($i, j \leq N$), by finding the location of the maximum of their cross-correlation [28]:

$$\Delta \nu = \underset{\nu}{\arg \max} \{ I_i(\nu) * I_j(-\nu) \}.$$

These shifts, including self-comparisons ($i = j$), are stored in an $N \times N$ matrix, illustrated for a representative 114-sample dataset in Fig. 2a. As an example, two typical spectra are shown in

Fig. 2b. The shift of each spectrum relative to the average position is computed by summing its mutual shifts and normalizing the result by $N$. We then co-align all of the spectra to the member of the set which has the lowest deviation from the average position and perform an average. The result of co-aligning just two such spectra is shown in Fig. 2c, illustrating the precision available from this technique. Finally, the average spectrum is shifted so that its center position coincides with the exact average position for the entire dataset.

**HITRAN fitting and étalon removal**

The HITRAN 2012 database [29] and an online spectral simulation tool [30] provide line-by-line spectroscopic parameters of a wide range of atmospheric gases, enabling the generation of high resolution absorbance spectra for water vapor and methane, which can be transformed into transmittance spectra using the Beer-Lambert law. These spectra are convolved with a Gaussian instrument response function of adjustable full width at half-maximum (FWHM) to simulate the limiting resolution of the experimental spectra. A multi-parameter fit is carried out using the Matlab Nelder-Mead simplex algorithm, in which the fitting parameters are the mole-fraction concentrations of water vapor and methane, a rigid frequency shift, the FWHM of the Gaussian instrument response function and 15–20 floating points which define the smooth envelope of the incident OPO spectrum via a spline function. The following parameters were fixed: temperature, at 300 K; pressure, at 1 atm; path lengths at 20 cm for methane and 2.3 m (2.5 m) for ambient water vapor with (without) the methane cell present. In Fig. 3a–c the fitted envelope function is shown in red and is compared with the average DCS spectrum, which is shown in blue. The fitting results are shown in Fig. 3d–f and the corresponding residuals appear in the blue traces in Fig. 3g–i. In the 7–8-μm band, parasitic étalon resonances occurring in the OPGaP crystals, a Ge filter and the ZnSe output couplers are revealed as sinusoidal oscillations in the residuals. The use of a fitting procedure which is both broadband and fits many parameters simultaneously is found to be extremely robust in the presence of these étalon effects. This is demonstrated by the immediate improvement in the fitting error after removing the étalon effects, as shown in Fig. 3j–l and the accompanying residuals (green) in Fig. 3g–i. Line shapes and features which are distorted or hidden before étalon-fringe removal are clearly resolved after the procedure is performed.

Étalon-fringe removal is implemented by Fourier transforming the original residuals to identify the principal frequencies present. As illustrated in Fig. 4—which takes as an example residuals from the water-vapor measurement of Fig. 3g—these frequencies can be presented on a scale calibrated in terms of the étalon optical thickness ($nL$), showing the four strongest étalons to have optical thicknesses of 3.18 mm, 8.36 mm, 0.45 mm and 7.23 mm. The first is attributed to the 1-mm OPGaP crystals ($n$ = 3.1780); the second is due to a 2-mm Ge filter ($n$ = 4.0058); the fourth is from the 3-mm ZnSe OPO cavity mirrors ($n$ = 2.4087). The exact origin of the third etalon is not certain but is very likely attributable to a thin ZnSe window in the HgCdTe detector. Using a peak detection algorithm, the principal étalon resonances can be isolated, allowing their contributions to be subtracted from the OPO transmission spectrum. Once this is done, the agreement with the HITRAN simulation is significantly improved, as shown in Fig. 3j–l.

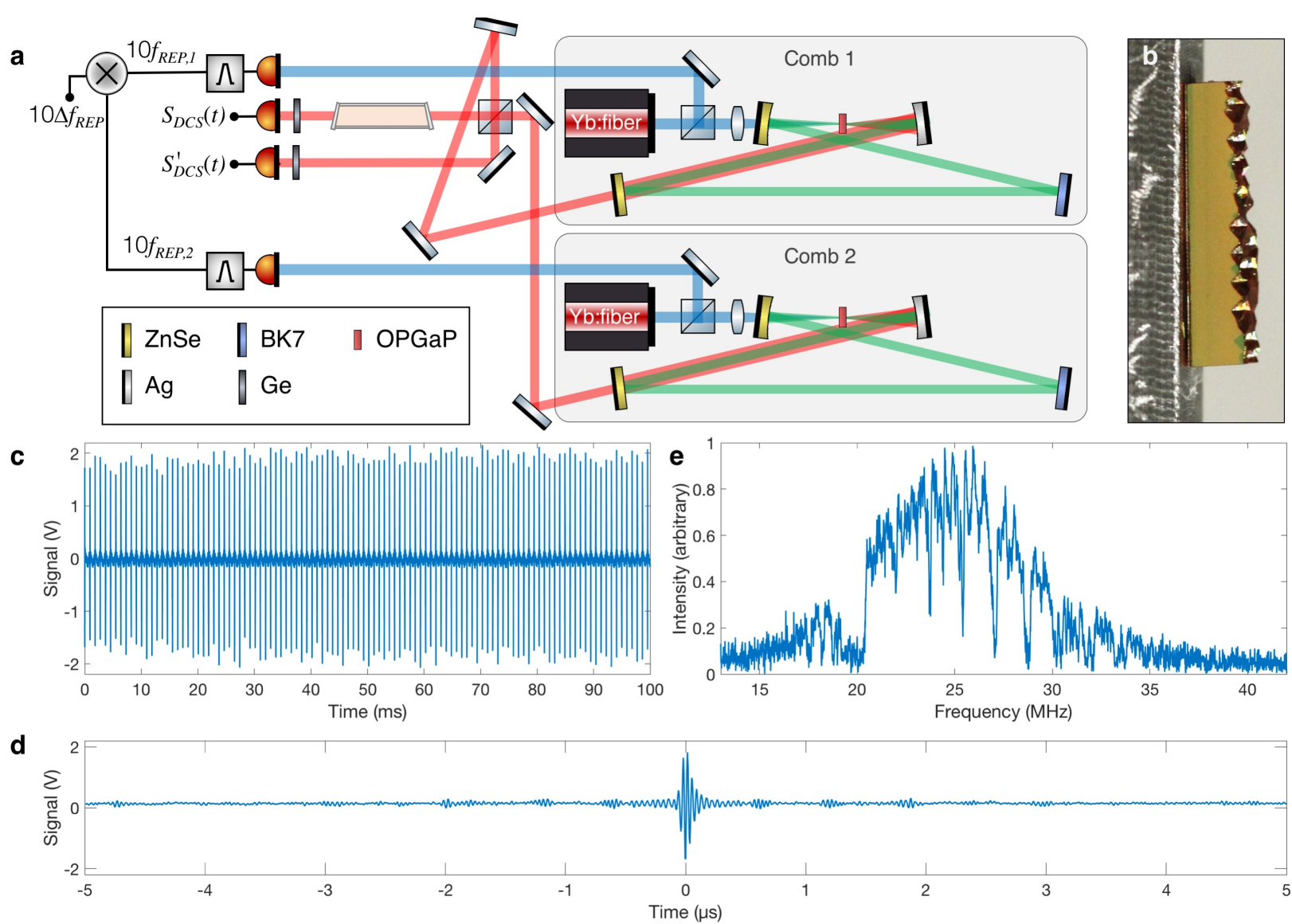

Fig. 1. Dual-comb spectroscopy using OPGaP optical parametric oscillators. **a.** Femtosecond pulses from two amplified Yb:fiber lasers of nearly identical repetition rates are used to synchronously pump two OPGaP OPOs. Idler pulses in the 6–8-μm range are output coupled from each OPO after reflection from a silver cavity mirror and transmission through a dielectric mirror on a ZnSe substrate. Idler pulses from each OPO are combined on a CaF$_2$ beamsplitter before detection by high speed HgCdTe detectors, one of which is preceded by a 20-cm CH$_4$ gas cell. **b.** Photograph of OPGaP crystal, showing the entrance facet. **c.** Example DCS dataset, comprising >100 interferograms acquired at ~1.1 kHz over 100 ms. The apparent intensity fluctuations are due to the different carrier-envelope phases of each interferogram. **d.** Detail of a single interferogram and **e.** its Fourier transform.

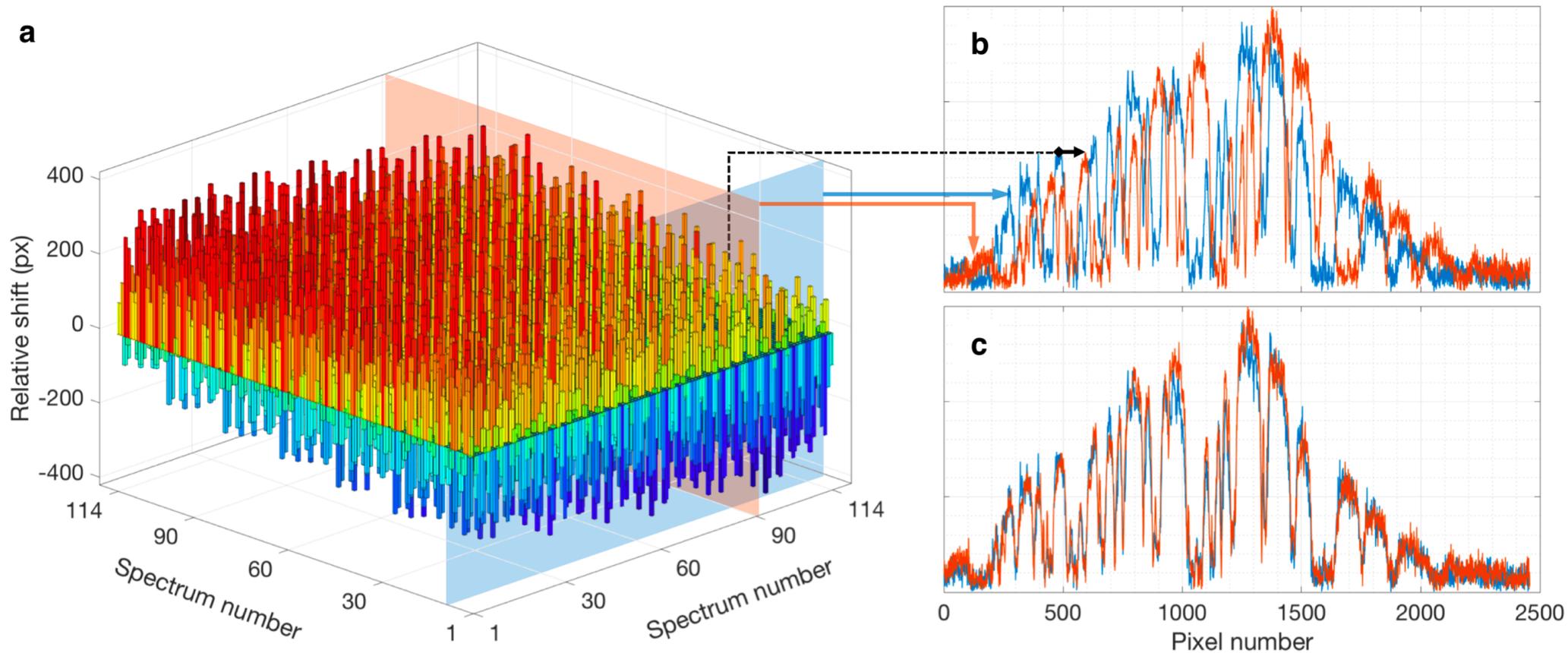

Fig. 2. Cross-correlation protocol for co-aligning multiple DCS spectra. **a.** Each member of a dataset of >100 spectra is cross-correlated individually with every other member to obtain a shift, expressed as a number of pixels and which can be visualized as an $N \times N$ matrix, where $N$ is the original number of DCS spectra. These shifts can be used to co-align the entire dataset, with **b.** and **c.** illustrating, respectively, an example of two such spectra before and after co-alignment.

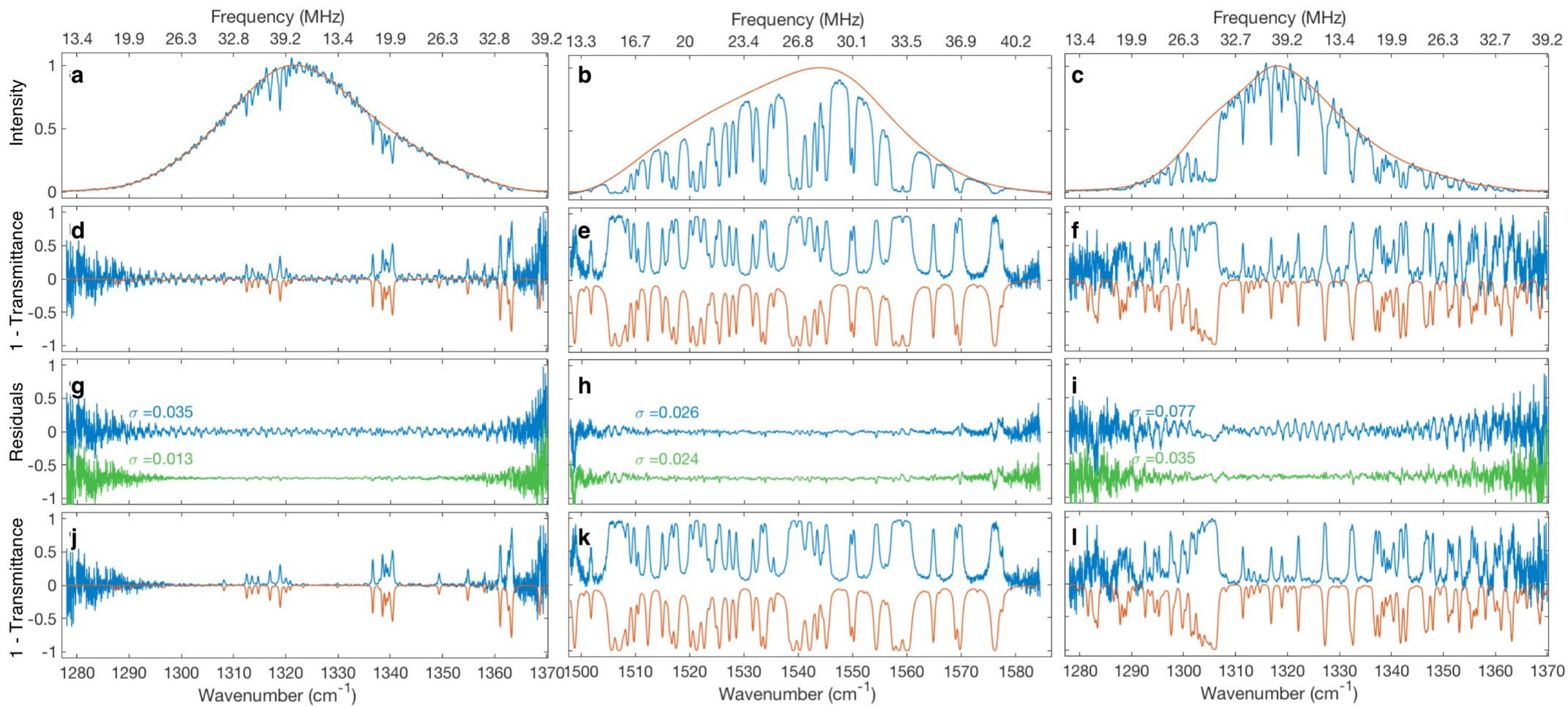

Fig. 3. Results of dual-comb fingerprint-region spectroscopy of ambient air (left and center) and with the additional of a 20-cm cell containing a nominal 2% methane in air mix (right). **a, b, c.** Averaged dual-comb spectra (blue) and fitted envelope (red); **d, e, f.** Extracted transmittance from the dual-comb data and comparison with a HITRAN 2012 fit, from which the spectroscopy resolution and gas concentrations are obtained. **g, h, i.** Residuals of the original fit (blue) and after étalon removal (green, vertically displaced by -0.6 for clarity). **j, k, l.** Transmittances after étalon removal, showing excellent agreement with the HITRAN 2012 lineshapes, magnitudes and positions.

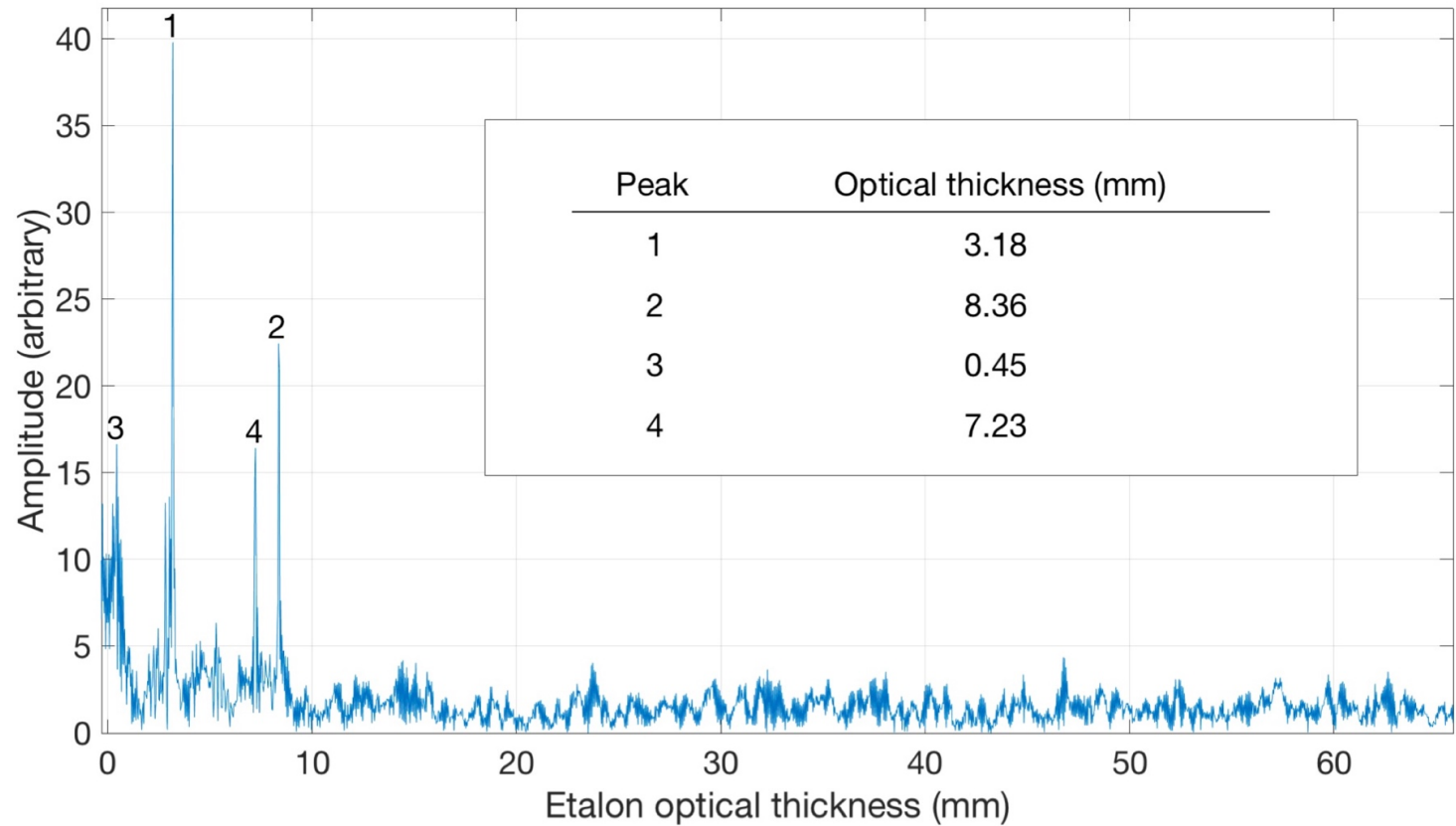

Fig. 4. Example of étalon removal protocol. Residuals obtained from the original full-spectrum fitting procedure (Fig. 3, d–f) are Fourier-transformed and the principal frequencies expressed in terms of étalon optical thickness. Peaks 1–4 are identified, respectively, with: the 1-mm OPGaP crystals (n = 3.1780); a 2-mm Ge filter (n = 4.0058); the 3-mm ZnSe OPO cavity mirrors (n = 2.4087); a thin ZnSe window in the HgCdTe detector (hypothesis).